\documentstyle[psfig]{mn}
\newcommand{\Carbon}{\mbox{${\rm {\rm ^{12}C}}$}}
\newcommand{\Nitrogen}{\mbox{${\rm {\rm ^{14}N}}$}}
\title{Erratum: Star formation and chemical evolution in SPH
simulations: a statistical approach}

\author[Lia, Portinari and Carraro]{ Cesario~Lia$^{1,2}$,  
Laura Portinari$^{3,2}$, 
and Giovanni Carraro$^2$ \\
       $^1$ SISSA/ISAS, via Beirut 4, I-34014 Trieste, Italy \\
       $^2$ Dipartimento di Astronomia, Universit\`a di Padova,
 Vicolo dell'Osservatorio 2, I-35122 Padova, Italy \\
       $^3$ Theoretical Astrophysics Center, Juliane Maries Vej 30,
 DK-2100 Copenhagen \O \\
E-mail: {\tt lportina@tac.dk, carraro@pd.astro.it}
}

\date{\tt Submitted: June 2002}

\pubyear{2002}
\begin{document}
\maketitle
\title{Erratum: SF and chemical evolution in SPH simulations}
%
%
\begin{keywords}
errata, addenda --- hydrodynamics --- methods: numerical --- stars: 
formation --- galaxies: evolution.
\end{keywords}


\noindent
The paper ``Star formation and chemical evolution in SPH
simulations: a statistical approach'' was published
on {\mbox{MNRAS}} 330, 821--836 (2002, astro-ph/0111084).
Here we notify a notation error in Equation~7 and 
a few typos in Table~3.

Eq.~7 on p.~826 should read as follows:
\[  N_{SN} \longrightarrow \frac{N_{SN}}{\int_t^{t+\Delta t} e(t') \, dt'} \]
with $e(t)$ given by Eq.~2 on p.~823. In fact, the particles that return 
to be gas within $\Delta t$ represent
a fraction $\int_t^{t+\Delta t} e(t') \, dt'$ of the whole parent stellar
population, although they are assigned a return probability 
$g_t(\Delta t)$
(Eq.~4) to consistently account for the star particles `lost' from the
initial population. Only as corrected above, Eq.~7 is consistent with
the corresponding metal production in Eq.~9 on p.~828. 

The mistake in Eq.~7 in the paper is the 
result of a different layout and notation of Eqs.~3,4 between
an early version of the manuscript and the final text. It is only
an error of notation and in our algorithm we applied the correct expression
presented here above, to obtain the correct SN rates shown e.g.\
in Fig.~5 and Fig.~9.

\medskip
\hspace{0truecm}
\begin{minipage}{18truecm}
\begin{small}
\begin{tabular}{|r r|c|c|c|c|}
\hline
\multicolumn{2}{|c|}{chemical element} & Salpeter & Kroupa & Arimoto--Yoshii & 
time range  [yr] \\
\hline
\hline
\Carbon & $p_C(t) = \left\{ 
\begin{array}{l}
 \\
 \\
 \\
\end{array} \right.$ & 
$\begin{array}{c}
1.2e{-10} \\
1.2e4~~t^{-2} \\
1.1e{-6}~~t^{-0.7} - 1.5e{-13}
\end{array} $ & 
$\begin{array}{c}
5.8e{-11} \\
5.8e3~~t^{-2} \\
2.9e{-8}~~t^{-0.5} - 4e{-13}
\end{array} $ & 
$\begin{array}{c}
3.6e{-10} \\
3.6e4~~t^{-2} \\
6.4e{-5}~~t^{-0.9} - 1e{-13}
\end{array} $ & 
$ \begin{array}{l}
t \in [3.4 \times 10^6, 10^7] \\
t \in [10^7, 3.4 \times 10^7] \\
t \in [2 \times 10^8, 5 \times 10^9] \\
\end{array} $  \\
 & & & & \\
\Nitrogen & $p_{Ns}(t)=$ & $7.7~Z_0~~t^{-1.4}$ & 
$0.98~Z_0~~t^{-1.3}$ & $520~Z_0~~t^{-1.6}$ & 
$t \in [3.4 \times 10^6, 15 \times 10^9]$ \\
\hline
\end{tabular}
\end{small}
\end{minipage}

\newpage
Table~3 on p.~827 contains one wrong entry (due to typsetting errors 
of the authors), affecting the yields of carbon 
$p_C(t)$ in the time range {\mbox{$[2 \times 10^8, 5 \times 10^9]$~yr}},
and a typo in the Kroupa entry for secondary nitrogen.
We repeat here below the complete entries for the yields of carbon
and secondary nitrogen, corrected where relevant.
\end{document}